\documentclass[prb,twocolumn,showpacs]{revtex4-1}
\usepackage{epsfig}
\usepackage{times}
\usepackage{graphicx}
\usepackage{amsfonts,amssymb}
\usepackage{amsmath}
\usepackage{color}
\definecolor{refcolor}{rgb}{1.0,0.0,0.0}

\newcommand{\be}{\begin{equation}}
\newcommand{\ee}{\end{equation}}   
\newcommand{\bea}{\begin{eqnarray}}
\newcommand{\eea}{\end{eqnarray}}
\newcommand{\ba}{\begin{array}}
\newcommand{\ea}{\end{array}}

\newcommand{\q}{{\bf q}}
\renewcommand{\k}{{\bf k}}
\newcommand{\Q}{{\bf Q}}

\begin{document}

\title{Spin-wave excitations in the SDW state of doped iron pnictides}
\author{Dheeraj Kumar Singh}
\email{dheerajsingh@hri.res.in} 
 
\affiliation{Harish-Chandra Research Institute, Chhatnag Road, Jhunsi, Allahabad 211019, India}
\affiliation{Homi Bhabha National Institute, Training School Complex, Anushakti Nagar, Mumbai 400085, India}
\begin{abstract}
We investigate the spin-wave excitations in the spin-density wave state of doped iron pnictides within a 
five-orbital model. We find that the excitations along ($\pi, 0$)$\rightarrow$($\pi, \pi$)
are very sensitive to the doping whereas they do not exhibit a similar sensitivity along ($0, 0$) $\rightarrow$ 
($\pi, 0$). Secondly, anisotropy in the excitations around ($\pi, 0$) with an elliptical shape grows on moving
towards the hole-doped region for low energy, whereas it decreases for high-energy excitations on the contrary. Thirdly, 
spin-wave spectral weight shifts towards the low-energy region on moving away from zero doping. We find 
these features to be in qualitative agreement with the inelastic neutron-scattering measurements for the doped pnictides.
\end{abstract}
\maketitle
\newpage
\section{Introduction}
Iron pnictides exhibit a very rich temperature-doping phase diagram,
where doping of either electrons or holes 
suppresses the long-range collinear magnetic order giving way to the 
sign-changing $s$-wave superconductivity (SC). Magnetic order in some
of these materials is stabilized over a range of
dopings $x_e \lesssim$ 0.06 and $x_h \lesssim$ 0.3 in
the electron- and hole-doped regions, respectively.\cite{dai,avci}
Despite the competing SC and magnetic long-range 
order on further doping, superconducting state retains the
spin fluctuations responsible for the pairing in a manner similar to
that in the high-$T_c$ cuprates.\cite{wang,terashimaa}
Experimentally, the role of spin fluctuations in mediating SC can be 
probed with the help of the inelastic neutron scattering 
(INS) a powerful experimental tool.\cite{dai2}

Origin of magnetic order in these materials lies 
in the Fermi surface (FS) instability because of a good 
nesting present between the Fermi pockets. According to the 
angle-resolved photoemission spectroscopy (ARPES)
as well as the band structure calculations, 
FSs consist of concentric hole pockets around $\Gamma$ 
and elliptical electron pocket 
at X.\cite{mazin,singh,haule,yi,kondo,yi1,brouet,kordyuk} 
Nesting between these two sets of pockets leads to the 
($\pi$, 0) SDW state or collinear magnetic order. When
electrons or holes are doped, FSs can be modified in 
a significant manner thereby altering the nature of
nesting, and that can have a significant impact 
on the nature of the SDW state as well as on the spin
fluctuations responsible for the SC.\cite{wang1}

A remarkably high-energy scale of the excitations
have been observed using INS in the SDW state of the
parent compound, which are highly dispersive as well as
sharp.\cite{ewings,diallo,zhao1,harriger,ewings1} 
They can extend up to $\sim$200 meV corresponding 
to the zone-boundary modes $\q = (\pi, \pi)$. There exists an 
in-plane anisotropy,\cite{zhao1} which persists even 
in the nematic phase.\cite{lu} Excitations have 
also been studied extensively for various dopings 
though in the superconducting state. For instance, the spin excitation is
manifested as a resonance in the superconducting state
\cite{maier,zhang} because of it's dependence on the BCS coherence
factors for the wavevector equal to the nesting vector 
and the opposite signs of the superconducting gap on the electron 
and hole pockets. Spin-wave dispersion similar to that
of parent compound BaFe$_2$As$_2$ has been observed 
along the high-symmetry directions for the superconducting
Ba$_{0.67}$K$_{0.33}$Fe$_2$As$_2$ with hole
doping $x \approx 0.33$ though with a significant
zone-boundary softening.\cite{wang1,horigane} At the same time, magnetic-exchange 
coupling is reduced by $10\%$. High-energy spin 
excitations are suppressed and magnetic spectral
weight is shifted to low energies. The excitations are gapped below 
$\sim$ 50 meV for the electron doped BaFe$_{1.7}$Ni$_{0.3}$As$_2$ while those of
high-energy are largely unaffected. It has
been further suggested that both the low-
and high-energy excitations may be associated 
with the superconductivity.\cite{wang1} 

For the SDW state, various theoretical and 
experimental studies have focused largely on the spin 
excitations in the parent compound. Two 
different types of itinerant models excitonic\cite{brydon} and
orbital\cite{kaneshita,nimisha,knolle,kovacic} have often been employed to understand the 
spin-wave excitations as well their
damping. The description of various 
characteristics of excitations has been
challenging in view of the 
fact that the observed magnetic moments
are small while the excitations are sharp 
and dispersing upto 200meV. Nonetheless, several 
characteristics have been captured within 
the five-orbital models.\cite{kaneshita,kovacic} In particular,
it has been shown recently that 
a large Hund's coupling plays an essential role
in describing different features such as sharpness,
anisotropy around X, and spin-wave spectral 
function.\cite{dheeraj} Another important factor that has a significant 
impact on the aforementioned features is the
doping induced modification in the bandstructure, which has
not attracted much attention.

In this paper, we examine various aspects of
the spin-wave excitations in the ($\pi, 0$) SDW 
state of doped iron pnictides within the
doping range $-0.4 \lesssim x \lesssim 0.05$. For this, we consider the 
five-orbital tight-binding model of Ikeda \textit{et al}.\cite{ikeda} 
The model with a rigid bandshift is known to exhibit ($\pi, 0$) SDW state within the 
doping range $0 \lesssim x_h \lesssim 0.18$ for 
intraorbital Coulomb interaction $U \sim 0.9$eV,\cite{schmiedt}
and the range can be expected to increase 
for relatively larger interaction parameters. 
The reconstructed FSs agrees 
qualitatively with the ARPES measurements.\cite{yi2} 
The quasiparticle interference obtained with the reconstructed 
bands within the model has reproduced several
features of the local density of state modulation 
in the doped state.\cite{chuang,dheeraj2} We use this model to
highlight three important consequences of doping on the spin-wave 
excitations. (i) The excitations along ($\pi, 0$)$\rightarrow$($\pi, \pi$) are very sensitive to the 
doping and a similar sensitivity is absent along
$(0, 0)$$\rightarrow$($\pi, 0$). (ii) Anisotropy around ($\pi, 0$) in the
form of elliptical structure increases on moving
towards the hole-doped region for the low-energy excitations, whereas 
it decreases for high-energy excitations on the
contrary. (iii) Spin-wave spectral weight shifts 
towards the low-energy region on moving away from zero doping.
\begin{figure}
\begin{center}
\vspace*{-4mm}
\hspace*{0mm}
\psfig{figure=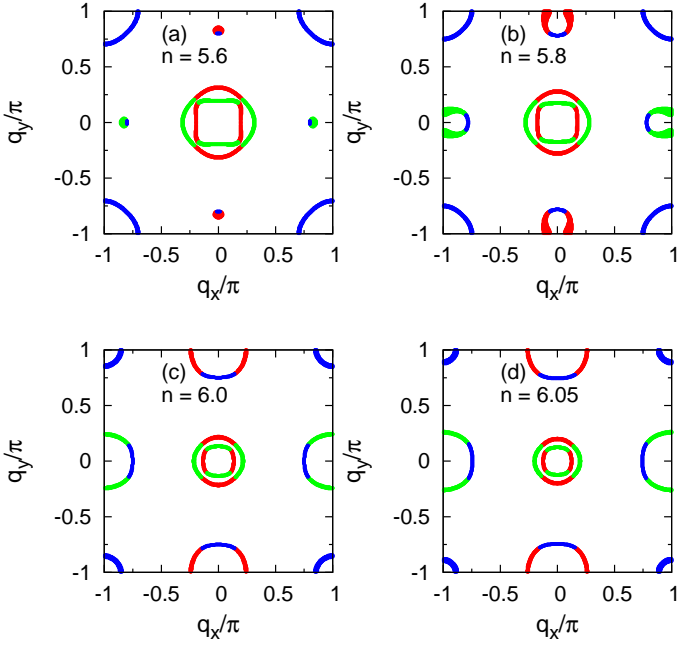,width=90.0mm,angle=0}
\vspace*{-12mm}
\end{center}
\caption{FSs in the unordered state calculated 
for the electron occupancies $n$ = (a) 5.6, (b) 5.8, (c) 6.0, and (d) 6.05.}
\label{fs1}
\end{figure}  
\begin{figure}
\begin{center}
\vspace*{-4mm}
\hspace*{0mm}
\psfig{figure=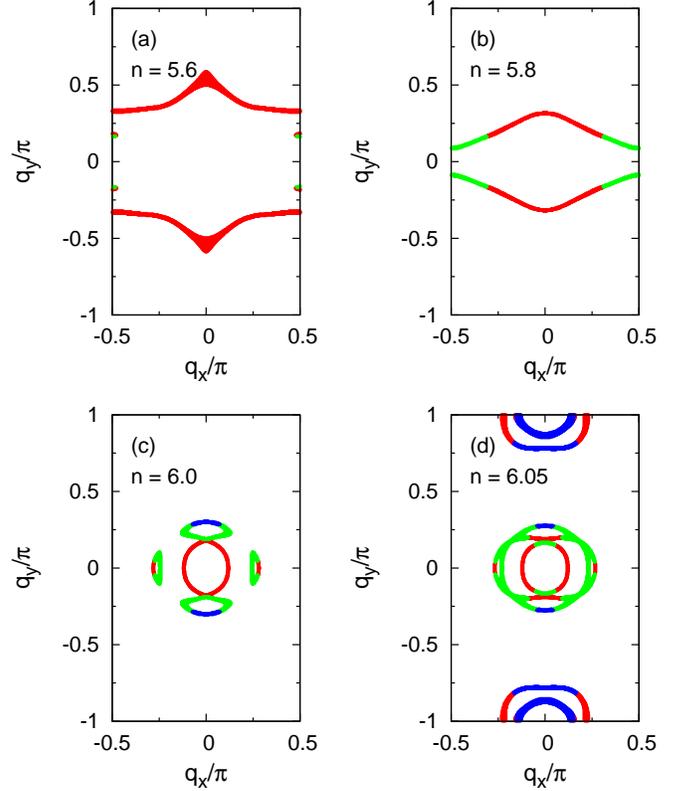,width=90.0mm,angle=0}
\vspace*{-12mm}
\end{center}
\caption{FSs in the SDW state with $U = 1.1$,
$J = 0.25U$ and $n$ = (a) 5.6, (b) 5.8, (c) 6.0, and (d) 6.05.}
\label{fs2}
\end{figure}  
\begin{figure}
\begin{center}
\vspace*{-2mm}
\psfig{figure=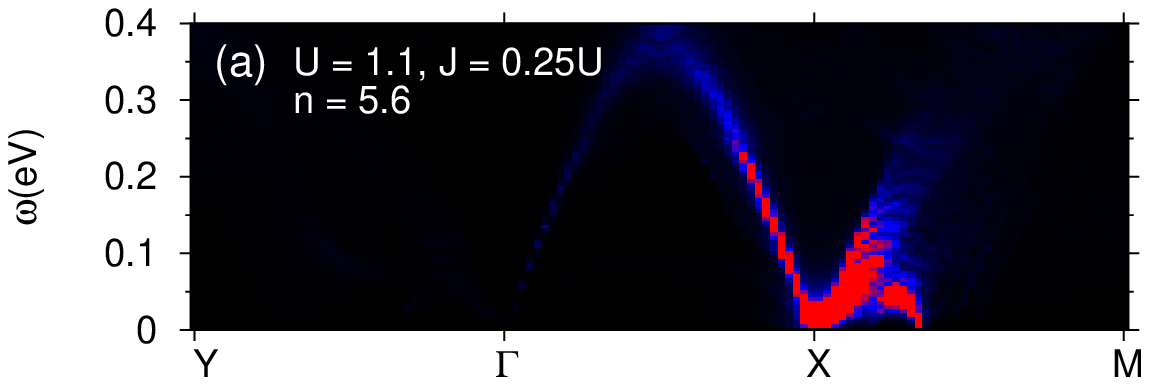,width=88mm,angle=0}
\vspace*{-0mm}
\psfig{figure=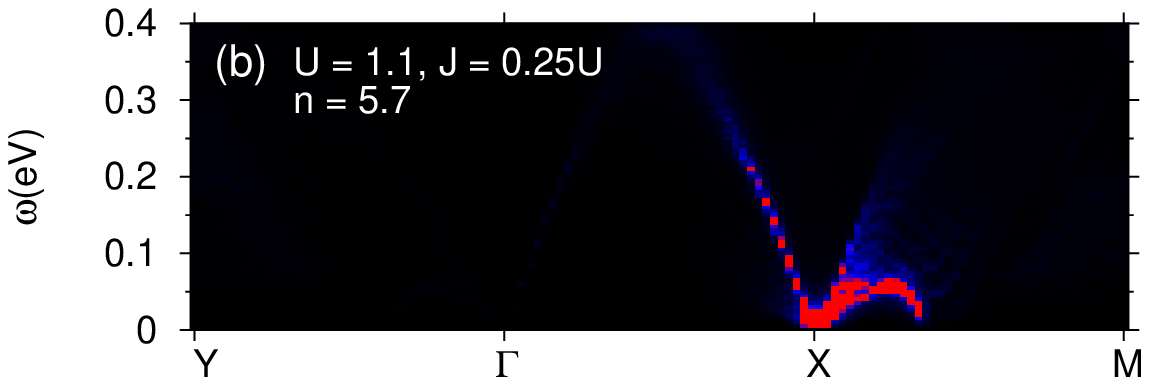,width=88mm,angle=0}
\vspace*{-0mm}
\psfig{figure=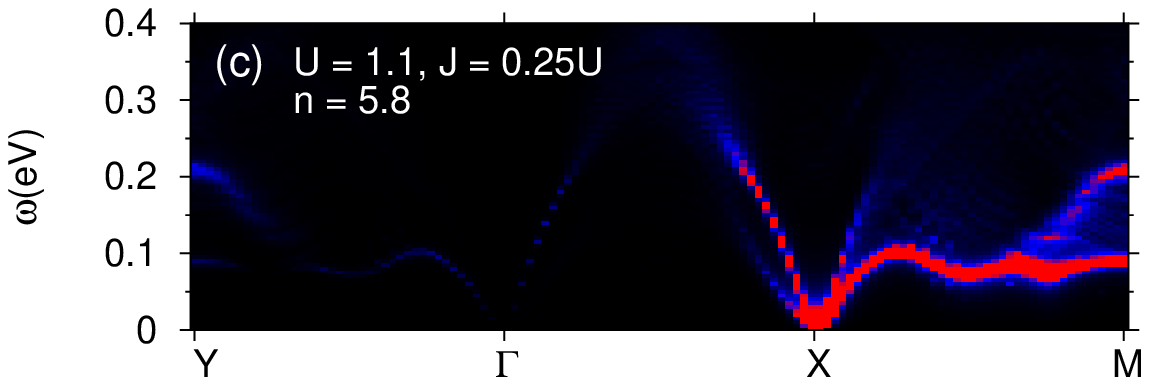,width=88mm,angle=0}
\vspace*{0mm}
\psfig{figure=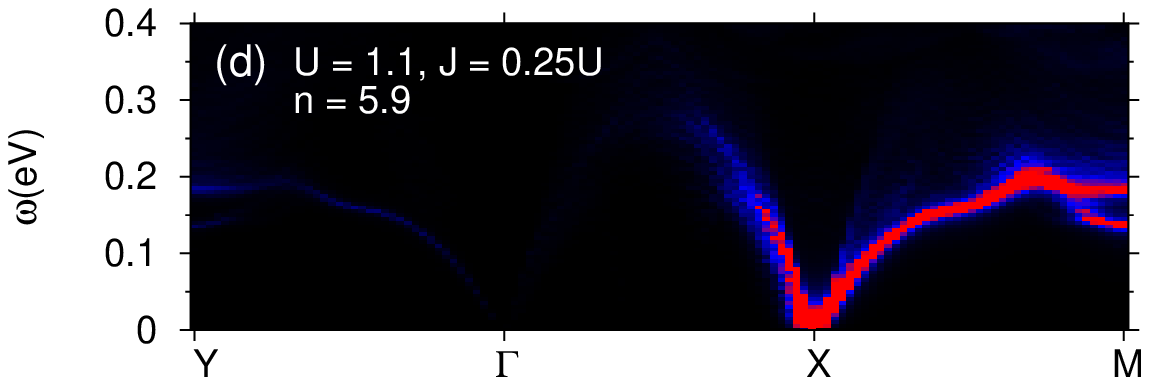,width=88mm,angle=0}
\vspace*{0mm}
\psfig{figure=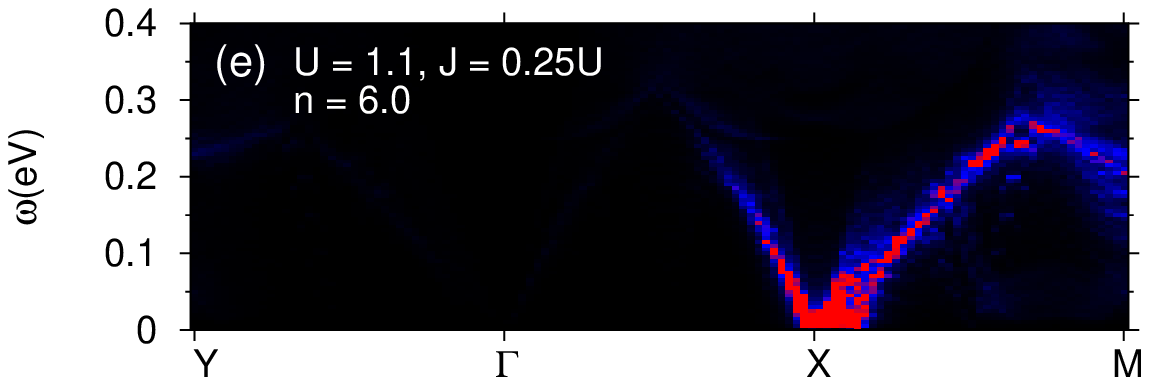,width=88mm,angle=0}
\vspace*{0mm}
\psfig{figure=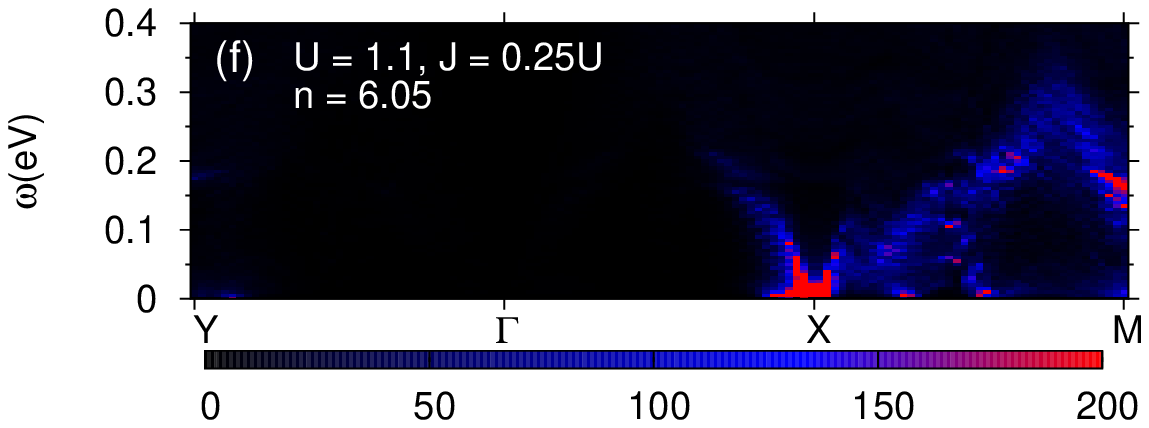,width=88mm,angle=0}
\end{center}
\vspace*{-0mm}
\caption{Imaginary part of the RPA-level physical transverse-spin susceptibility 
Im$\bar{\chi}^{ps}_{\rm RPA}(\q, \omega)$ calculated for $U = 1.1$, $J = 0.25U$ and several 
values of $n$ = (a) 5.6, (b) 5.7, (c) 5.8, (d) 5.9,
(e) 6.0, and (f) 6.05.}
\label{rpa}
\end{figure}  
\begin{figure}
\begin{center}
\vspace*{-2mm}
\psfig{figure=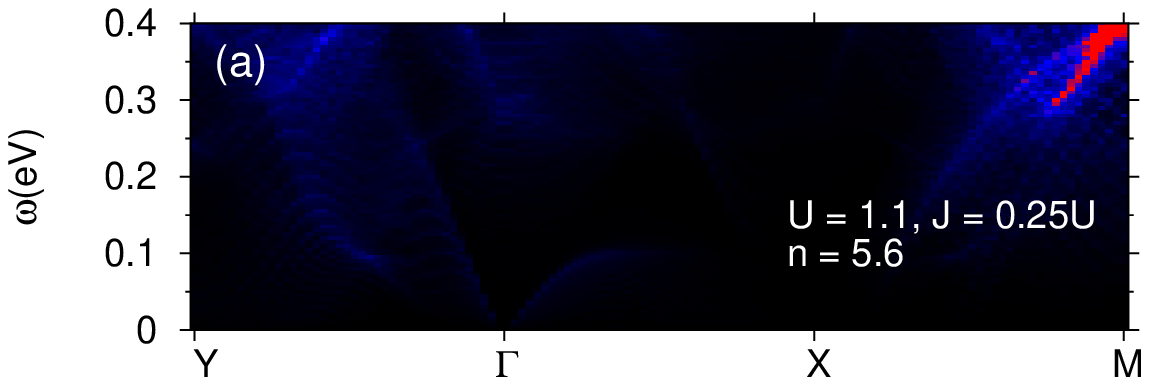,width=88mm,angle=0}
\vspace*{-0mm}
\psfig{figure=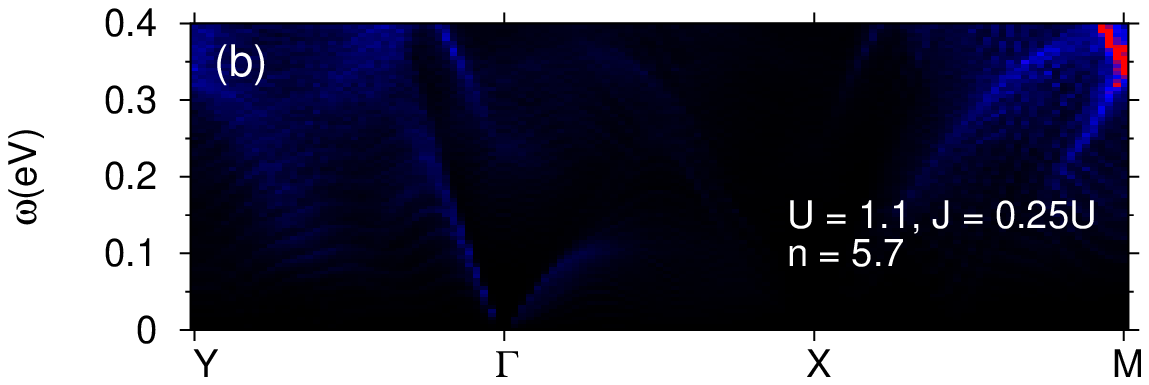,width=88mm,angle=0}
\vspace*{-0mm}
\psfig{figure=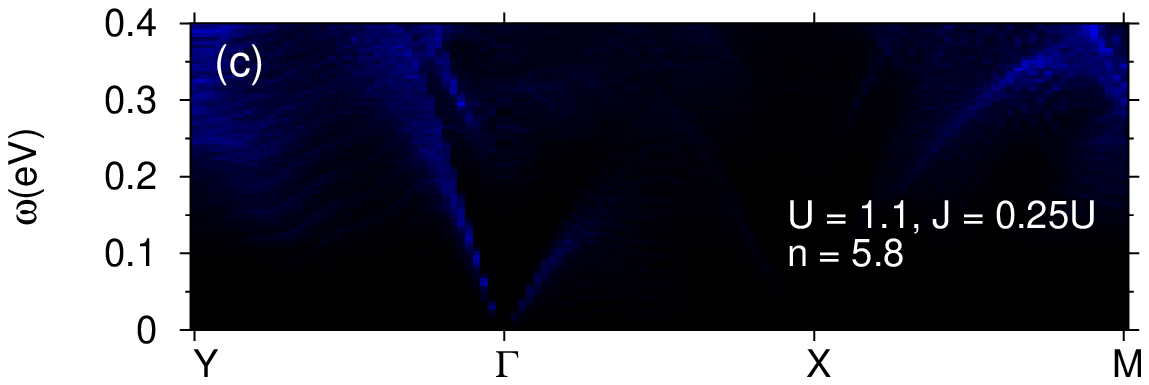,width=88mm,angle=0}
\vspace*{0mm}
\psfig{figure=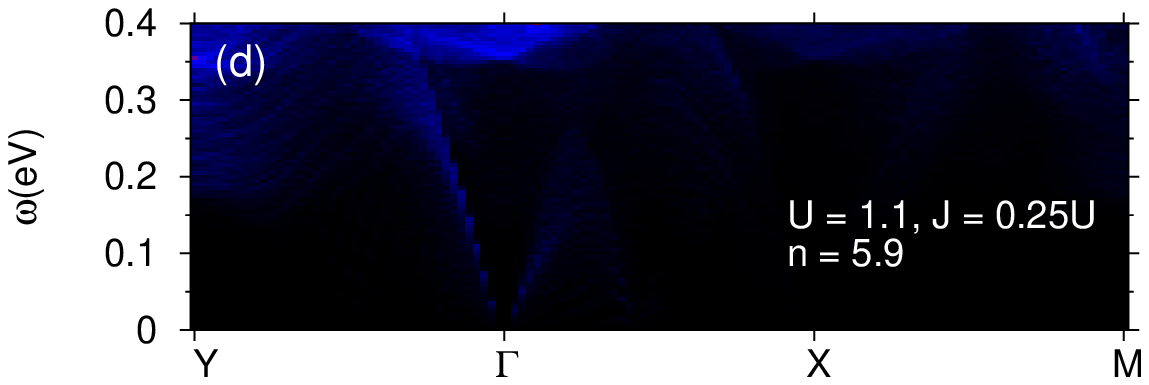,width=88mm,angle=0}
\vspace*{0mm}
\psfig{figure=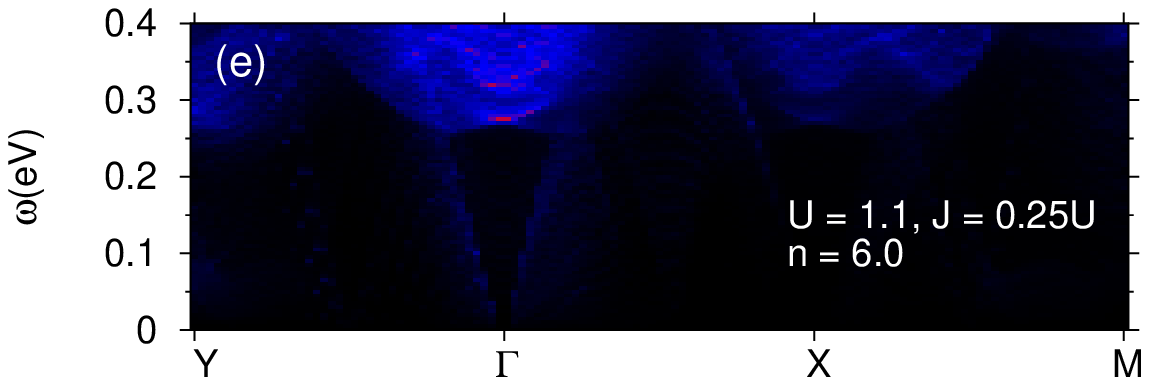,width=88mm,angle=0}
\vspace*{0mm}
\psfig{figure=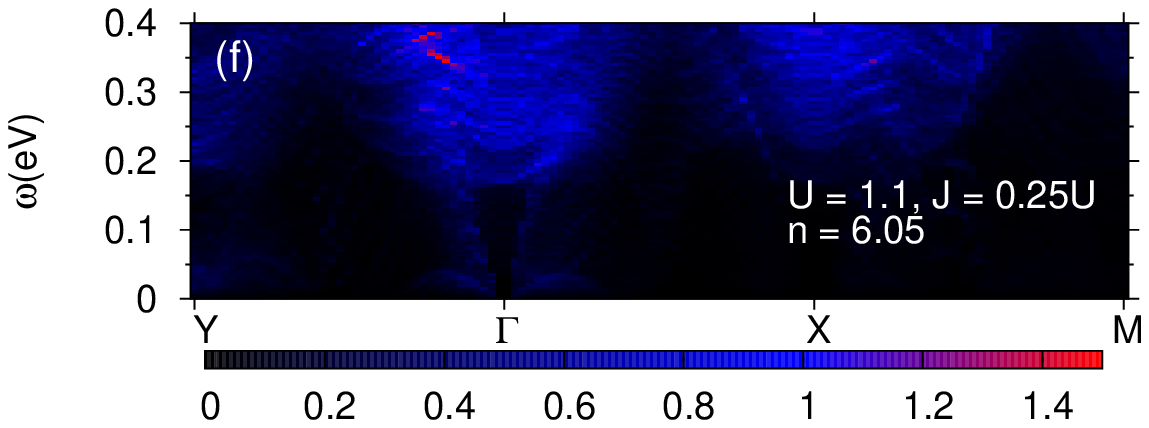,width=88mm,angle=0}
\end{center}
\vspace*{-0mm}
\caption{Imaginary part of the bare spin susceptibility Im$\bar{\chi}^{ps}_{\rm RPA}(\q, \omega)$
calculated for the parameters as in Fig. \ref{rpa}.}
\label{bare}
\end{figure}  
\begin{figure}
\begin{center}
\vspace*{-8mm}
\hspace*{-8mm}
\psfig{figure=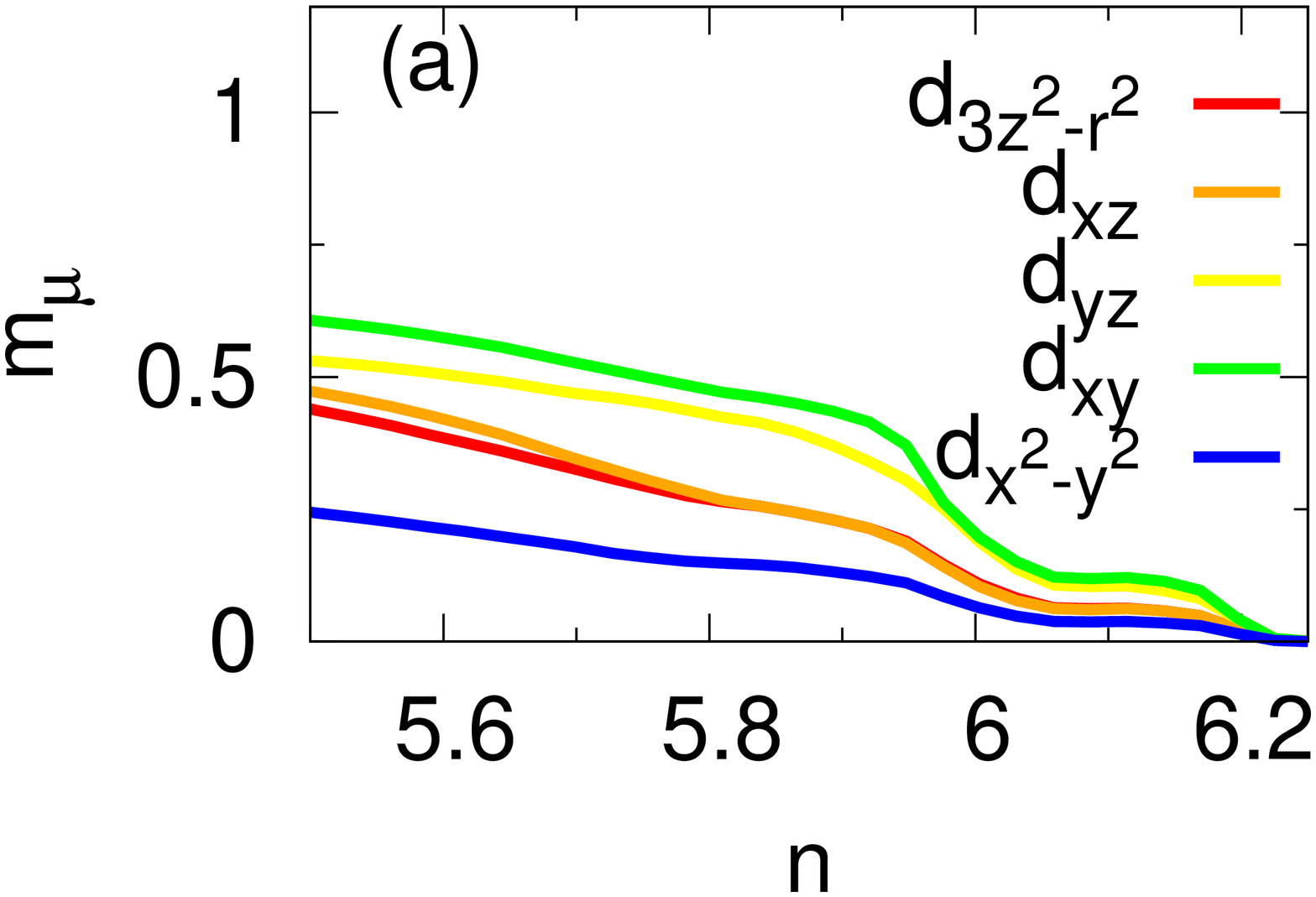,width=32.0mm,angle=-90}
\hspace*{-8mm}
\psfig{figure=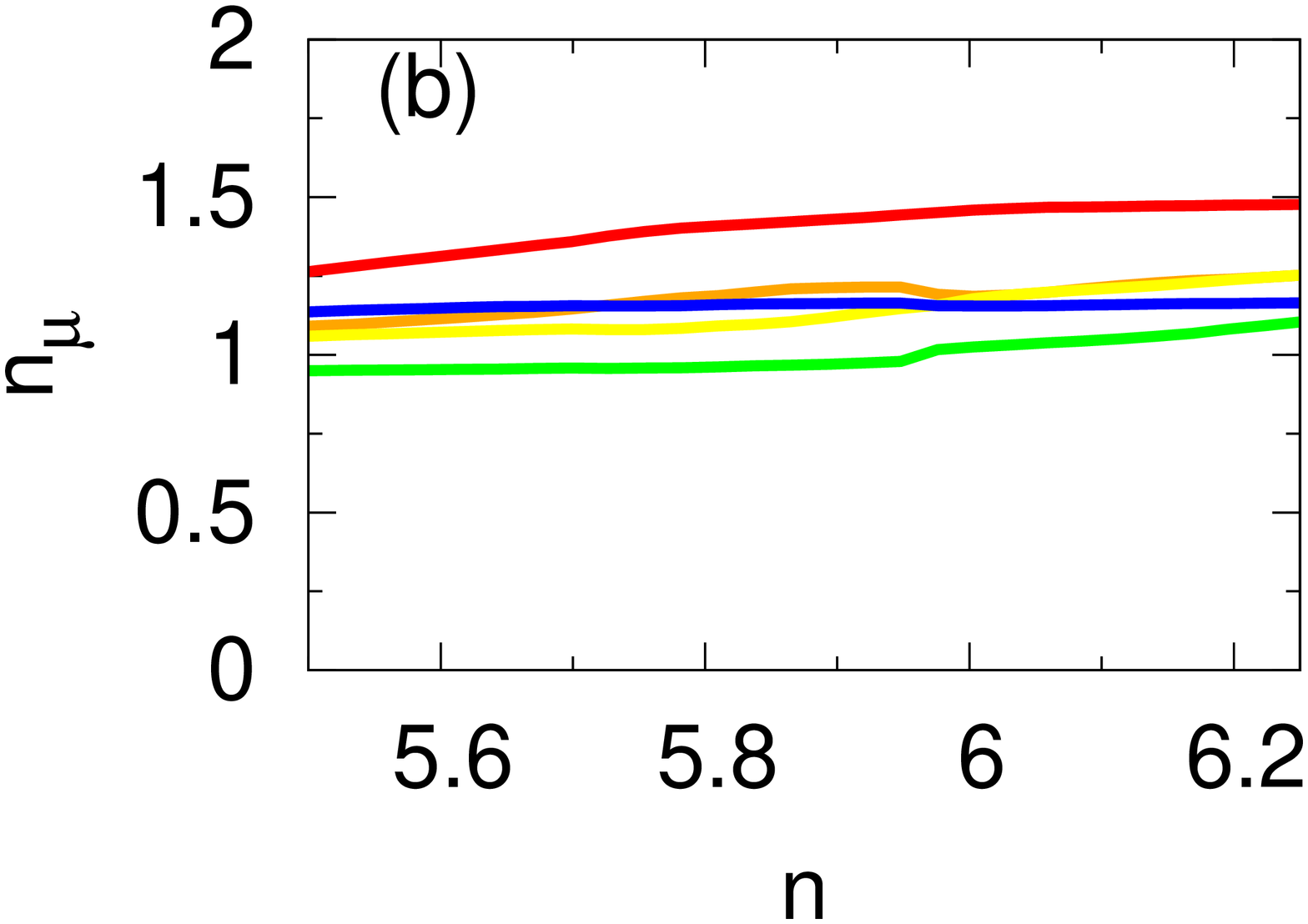,width=32.0mm,angle=-90}
\vspace*{-6mm}
\end{center}
\caption{Orbital-resolved (a) magnetizations and (b) charge densities as a function of electron and hole dopings.}
\label{}
\end{figure}  
\begin{figure}
\begin{center}
\vspace*{-6mm}
\hspace*{0mm}
\psfig{figure=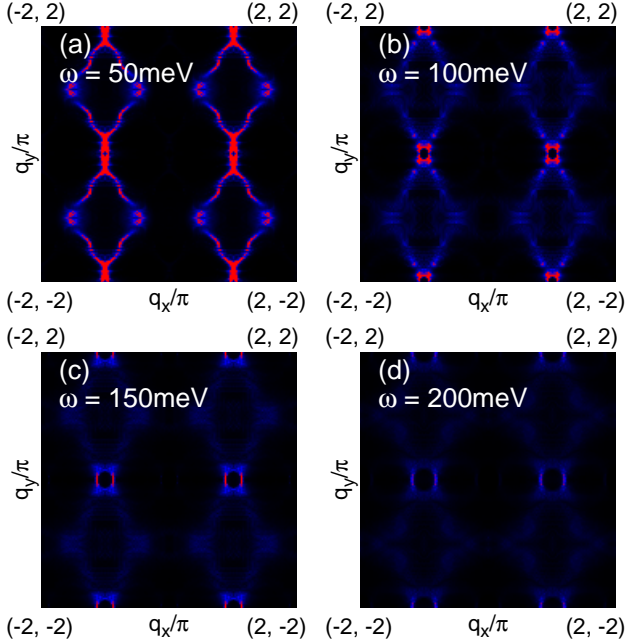,width=90.0mm,angle=0}
\vspace*{-12mm}
\end{center}
\caption{Constant energy cuts of Im$\bar{\chi}^{ps}_{\rm RPA}(\q, \omega)$ for $U = 1.1, J = 0.5U$ and $n = 5.7$ at
energies (a) 50 meV, (b) 100 meV, (c) 150 meV, and (d) 200 meV.}
\label{cut1}
\end{figure}  
\begin{figure}
\begin{center}
\vspace*{-6mm}
\hspace*{0mm}
\psfig{figure=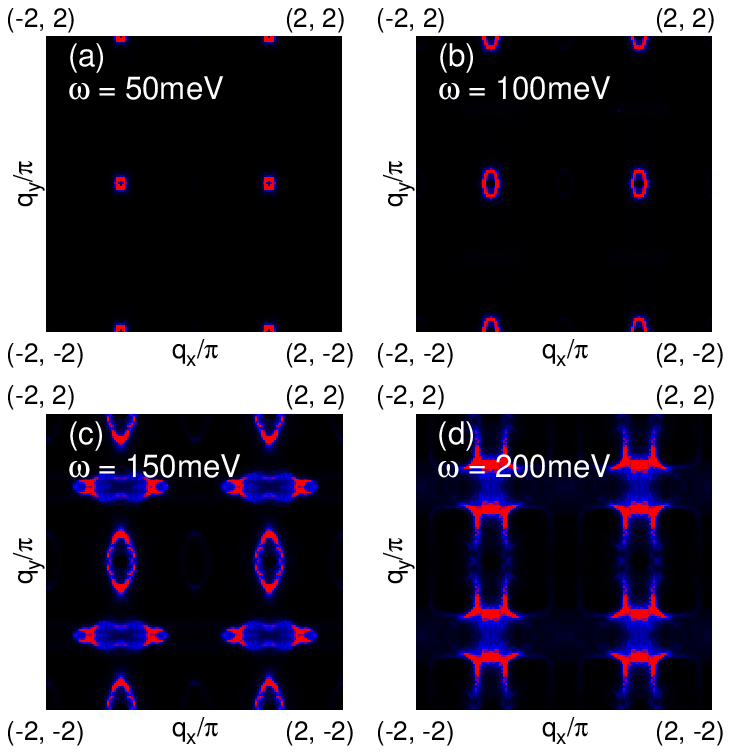,width=90.0mm,angle=0}
\vspace*{-12mm}
\end{center}
\caption{Constant energy cuts of Im$\bar{\chi}^{ps}_{\rm RPA}(\q, \omega)$
for $n = 5.9$ with other parameters as in Fig. \ref{cut1}.}
\label{cut2}
\end{figure}  
\begin{figure}
\begin{center}
\vspace*{-6mm}
\hspace*{0mm}
\psfig{figure=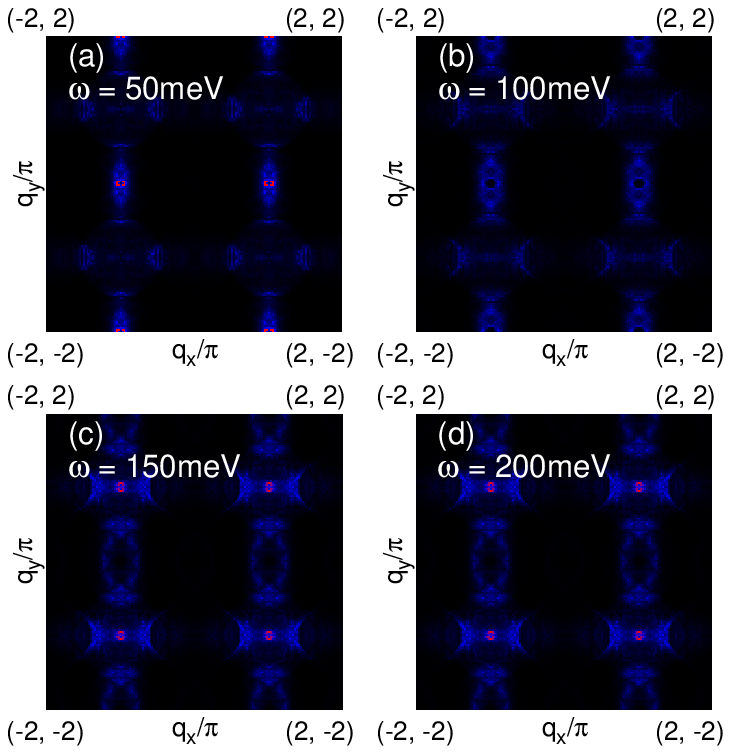,width=90.0mm,angle=0}
\vspace*{-12mm}
\end{center}
\caption{Constant energy cuts of Im$\bar{\chi}^{ps}_{\rm RPA}(\q, \omega)$ for $n = 6.05$
with other parameters as in Fig. \ref{cut1}.}
\label{cut3}
\end{figure}  
\section{Transverse-spin fluctuations}
In order to investigate the doping dependence of spin-wave excitations, 
we consider the following mean-field Hamiltonian in the ($\pi, 0$) SDW state 
\be
\hat{\mathcal{H}}_{{\sigma}\k} = 
\sum_{\k \sigma}\Psi^{\dagger}_{{\bf k} \sigma}
\begin{pmatrix}
 \hat{\varepsilon}_{\k}+\hat{N} \,& \,{\rm sgn}\bar{\sigma}\hat{\Delta} \\
 {\rm sgn}\bar{\sigma}\hat{\Delta} \,& \,\hat{\varepsilon}_{\bf {k+Q}}+\hat{N}
\end{pmatrix} 
\Psi_{{\bf k} \sigma}.
\ee
Here, $\Psi^{\dagger}_{{\bf k} \sigma} = (d^{\dagger}_{{\k}1\sigma},....,d^{\dagger}_{{\k}5\sigma},
{d}^{\dagger}_{{\k}\bar{1}\sigma},....,{d}^{\dagger}_{\k\bar{5}\sigma})$ with $\bar{d}^{\dagger}_{{\k}\bar{\mu}\sigma}$ = 
$d^{\dagger}_{{\k+\Q}\mu\sigma}$. $d^{\dagger}_{{\k} i \sigma}$ ($d_{{\k}1\sigma}$) is the electron 
creation (destruction) operator for the momentum ${\k}$ in the orbital $i$ with spin $\sigma$. $\hat{\varepsilon}_{\k}$ is the hopping matrix corresponding to the 
five-orbital model. The elements of
the matrices $\hat{\Delta}$ and $\hat{N}$ as described in 
the Appendix are dependent on the onsite interaction
parameters, orbital magnetizations, and charge 
densities. The Hamiltonian matrix is diagonalized and 
various order parameters are obtained in a self-consistent manner
using the eigenvalues and eigenvectors. Then, 
the eigenvalues and eigenvectors corresponding to the self-consistent SDW state 
are used to calculate the spin-wave excitations.

The transverse-spin susceptibility within the
random-phase approximation can be obtained as
\be
\hat{\bar{\chi}}_{\rm RPA}(\q, i\omega_n) = (\hat{{\bf 1}} -
\hat{U})^{-1} \hat{\bar{\chi}}(\q, i\omega_n),
\ee
where $\hat{{\bf 1}}$ is a $2n^2 \times 2n^2$ identity
matrix with $n = 5$. The elements of block-diagonal matrix 
${U}_{\mu_1 \mu_2;\mu_3 \mu_4}$ = $U,\,U-2J,\,J$ and $J$ for $\mu_1=\mu_2=\mu_3=\mu_4,\,
\mu_1=\mu_2\ne\mu_3=\mu_4 ,\,\mu_1 = \mu_2\ne \mu_3 =
\mu_4$ and $\mu_1=\mu_4\ne \mu_2=\mu_3$,
respectively. They vanish otherwise. $U$ and $J$ are the intraorbital and 
interorbital Coulomb interactions. The bare-level susceptibility matrix is 
\be
\hat{\bar{\chi}}(\q,i\omega_n) = 
\begin{pmatrix}
 \hat{\bar{\chi}}(\q,\q,i\omega_n) \,& \,\hat{\bar{\chi}}(\q,\q+\Q,i\omega_n) \\
 \hat{\bar{\chi}}(\q+\Q,\q,i\omega_n) \,& \,\hat{\bar{\chi}}(\q+\Q,\q+\Q,i\omega_n)
\end{pmatrix},
\ee
where the elements in the ordered state is
given by
\bea
\bar{\chi}^{}_{\alpha \beta, \mu \nu} = \chi^{+-}_{\alpha \beta, \mu \nu}
+\chi^{+-}_{\bar{\alpha} \beta, \bar{\mu} \nu}+\chi^{+-}_{\alpha \bar{\beta}, \mu \bar{\nu}}
+\chi^{+-}_{\bar{\alpha} \bar{\beta}, \bar{\mu} \bar{\nu}},
\eea
where Umklapp processes are included.
Physical transverse-spin susceptibility corresponding to 
the spin operators to be defined below is  
\be
\bar{\chi}^{ps}(\q, i\omega_n) = \sum_{\alpha \mu} {\bar{\chi}}_{\alpha \alpha,
\mu \mu} (\q, \q, i\omega_n).
\ee

The transverse-spin susceptibility is defined as  
\bea
\chi^{+-}_{\alpha \beta, \mu \nu}(\q,\q^{\prime},i\omega_n)  \nonumber\\
&=&\frac{1}{\beta} \int^{\beta}_0{d\tau e^{i \omega_{n}\tau}\langle T_\tau
[{ S}^{+}_{\alpha \beta}(\q, \tau) {S}^{-}_{\nu \mu} (-\q^{\prime}, 0)]\rangle}.
\eea
where components of the spin operator are 
\be
{\cal S}^{i}_{\q}= \sum_{\bf k} \sum_{\sigma \sigma^{\prime}}
\sum_{\mu \mu^{\prime}} d^{\dagger}_{\mu \sigma}(\k+\q)
E_{\mu \mu^{\prime}} \sigma^{i}_{\sigma \sigma^{\prime}} d_{\mu^{\prime} \sigma^{\prime}}(\k).
\ee
$E$ is a 5$\times$5 identity matrix belonging to the orbital 
bases. Thus the susceptibility when $\q^{\prime} = \q$ is given 
in terms of Green's function as
\bea 
\chi^{+-}_{\alpha \beta, \mu \nu}(\q,\q,i\omega_n)  \nonumber\\
&=& \sum_{\k,i\omega^{\prime}_n} G^{0\uparrow}_{\alpha \mu}(\k+\q,i\omega^{\prime}_n
+i\omega_n)G^{0\downarrow}_{\nu \beta} (\k,i\omega^{\prime}_n).
\eea

In the following, analytic continuation $i\omega_n \rightarrow \omega + i
\eta$ with $\eta = 2$meV is used throughout. Interaction 
parameters $U$ and $J$ are set to be 1.1eV and 0.25$U$ so
as to obtain magnetic moment $m \sim 1$ for zero doping
motivated by the observed magnetic moments in 122 series 
of pnictides.\cite{huang} Unit of energy 
is set to be eV unless stated otherwise.
\section{results}
FSs obtained in the unordered state for various dopings 
(a) $n = 5.6$, (b) $n = 5.8$, (c) $n = 6.0$, and (d) $n = 6.05$ are 
shown in Fig. \ref{fs1}. As can be seen, the nesting deteriorates fast on moving 
away from zero doping because of the electron and hole pockets
shrink and expand on moving from the electron-doped to hole-doped region, respectively. Fig. \ref{fs2} shows 
the reconstructed FSs in the SDW state for the respective dopings. It is evident that the 
reconstructed structure and topology of the FSs are very sensitive to the
doping when compared to the unordered state.

Fig. \ref{rpa} shows Im$\bar{\chi}^{sp}_{\rm RPA}$ obtained in 
the SDW state for various dopings (a) 5.6, (b) 5.7, (c) 5.8 ,(d)
5.9, (e) 6.0, and (f) 6.05. Note that the red color also represents those value of Im$\bar{\chi}^{sp}_{\rm RPA}$ that 
exceed 200. The excitations are heavily 
damped throughout particularly 
along $\Gamma$-X-M ((0, 0)$\rightarrow$($\pi, 0$)$\rightarrow$($\pi, \pi$)) for the electron doping $n = 6.05$, 
which is not surprising because there is a fast reduction in
the net magnetization, and hence of the magnetic-exchange
gap on electron doping. That is reflected in the particle-hole 
continuum extending down to low energy in the bare 
susceptibility (Fig. \ref{bare}). The nesting between 
the electron pocket and the hole pocket in the unordered state is optimal
in the vicinity of $n \approx 6.0$. Therefore, moving
away from this band filling is expected to lead to a 
reduced magnetic-exchange gap according to the nesting-based
scenario though the reduction may not necessarily be 
symmetrical about zero doping.

Spin-wave damping reduces quickly on moving towards 
the hole-doped region and the 
excitations become optimally sharp and well-defined 
for $n \approx 5.9$ in the hole-doped region. In an earlier work, maximum $T_c$ for the SDW state
has also been shown to occur within the model 
for small hole doping. On doping holes further, the
spin-wave excitations get
softened rapidly along X-M especially in the region 
close to M. Meanwhile, damping increases so that the excitations disappear finally
for $n \approx 5.75$ in a large part of X-M. 
However, they remain largely unaffected along $\Gamma$-X.

Magnetic moment grows on moving from the electron-doped 
region to the hole-doped region as the system approaches half filling (Fig. 5). This is expected to 
suppress the density of states at the Fermi level and the electron movement in the lattice leading to the
departure from metallicity.\cite{medici,lafuerza} Despite that the FSs exist
for the all the dopings considered here because of 
a small ratio $U$/$W \approx 0.25$ with $W$ as the electron
bandwidth, and consequently reconstructed 
bandstructure plays a crucial role in the 
SDW excitations. For instance, the particle-hole 
continuum in the bare-spin susceptibility 
shifts towards low energy near
M, which should play a very important role 
in the damping and disappearance of
the spin-wave excitations along X-M in the hole-doped region. 

Figs. \ref{cut1}, \ref{cut2} and \ref{cut3} 
show the constant energy cuts for the bandfilling $n = 5.7, 5.9\,
\,{\rm and}\,\,6.05$, respectively. It is noted that the anisotropy in the spin-wave excitations around X 
is sensitive to both energy as well as doping. It 
diminishes on increasing the energy whereas
grows on moving from the electron-doped region to the
hole-doped region for the low-energy 
excitations $\sim$ 50meV. On the other hand, it 
decreases for the high-energy excitations
$\gtrsim$ 100meV. The anisotropic behavior 
around X at low energy as well as square like shape of excitations around M 
at higher energy particularly when $n \approx 5.9$ is qualitatively similar
to what is observed in the INS experiments.\cite{wang1,dai2}  
\begin{figure}
\begin{center}
\vspace*{-4mm}
\hspace*{0mm}
\psfig{figure=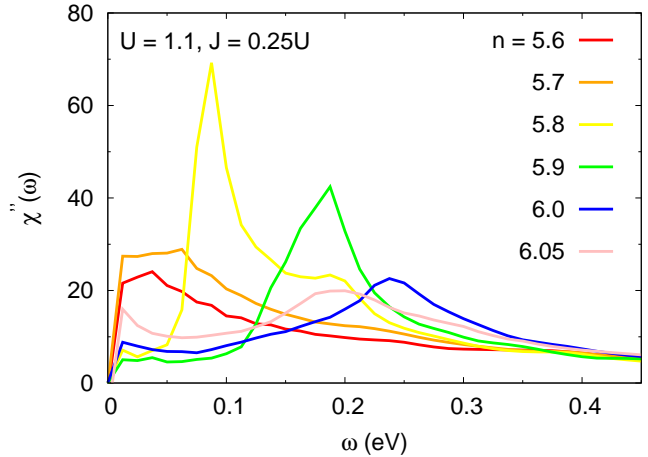,width=90.0mm,angle=0}
\vspace*{-10mm}
\end{center}
\caption{Spin-wave spectral function calculated for several dopings 
$n$ = 5.6, 5.7, 5.8, 5.9, 6.0, and 6.05.}
\label{spect}
\end{figure}  
 
Fig. \ref{spect} shows the doping dependence 
of the spin-wave spectral function. 
There are two separate peaks for $n = 6.05$, one positioned near very 
small energy and other one around 200meV. At zero doping, the
peak near 200meV gets prominent. Whereas the position of the peak 
shifts rapidly towards the low-energy region on hole doping. A similar 
observation has been made in the paramagnetic and superconducting phases of
doped pnictides. Moreover, the location of peak in the spin-wave spectral
function for the optimal doping is near 200meV, which is also
in accordance with the experiments.\cite{wang1}
\section{Conclusions and Discussions}
In conclusions, we have investigated the
spin-wave excitations in the SDW state of doped iron pnictides. 
We use a five-orbital model with realistic
electronic structure and fixed set of interaction parameters corresponding 
to the magnetic moment $1\mu_B$ in the 
undoped case. We find that the excitations along ($\pi, 0$)$\rightarrow$($\pi, \pi$) are very sharp and 
dispersive in a very small doping range
centered around $n \approx 5.9$ lying in the hole-doped 
region and they get damped heavily on moving away 
from that doping on either side. Unlike
along ($\pi, 0$)$\rightarrow$($\pi, \pi$), the excitations
along $(0, 0)$$\rightarrow$($\pi, 0$) do not show much variation with 
doping except in the electron-doped region, where they
are damped heavily. Further, 
we find that the anisotropy around X decreases at
high energy for any doping. Whereas it increases and 
decreases on hole doping for low- and high-energy
excitations, respectively. We also find that the spin-wave spectral weight 
shifts towards lower energy on doping either holes or electrons, whereas it is 
peaked near 200meV for the undoped case. 

In several doped pnictides, there is a phase 
transition from the high-temperature SDW state to the low-temperature 
superconducting state within a certain doping
range. Some of the characteristics of the spin excitations such as 
anisotropy around ($\pi, 0$) are expected to be retained 
across the phase transition in a manner 
similar to the phase transition from the paramagnetic to SDW state. 
We find several features such as 
anisotropy around ($\pi, 0$), softening of zone-boundary modes, 
and shifting of the spectral weight on doping to the low-energy 
region being in qualitatively similar to those measured in the INS for 
the paramagnetic and superconducting phases.

We acknowledge the use of HPC clusters at HRI.

\section*{Appendix}
The kinetic part of the model Hamiltonian that we consider
is given by
\begin{equation}
 \mathcal{H}_0 = \sum_{\k}\sum_{\mu,\nu}\sum_{\sigma} \varepsilon_{\k}^{\mu\nu} 
d_{\k \mu\sigma}^\dagger d_{\k \nu\sigma} + \text{H.c.}, 
\end{equation} 
where $d_{{\bf k} \mu \sigma}^\dagger$ ($d_{{\bf k} \mu \sigma}$) is the electron creation
(destruction) operators and $\varepsilon_{\k}^{\mu\nu}$ are the hopping elements from orbital 
$\mu$ to $\nu$ for the momentum $\k$. 

The interaction term is given by
\begin{eqnarray}
\mathcal{H}_{int} &=& U \sum_{{\bf i},\mu} n_{{\bf i}\mu \uparrow} n_{{\bf i}\mu \downarrow} + (U' -
\frac{J}{2}) \sum_{{\bf i}, \mu<\nu} n_{{\bf i} \mu} n_{{\bf i} \nu} \nonumber \\ 
&-& 2 J \sum_{{\bf i}, \mu<\nu} {\bf{S_{{\bf i} \mu}}} \cdot {\bf{S_{{\bf i} \nu}}} + J \sum_{{\bf i}, \mu<\nu, \sigma} 
d_{{\bf i} \mu \sigma}^{\dagger}d_{{\bf i} \mu \bar{\sigma}}^{\dagger}d_{{\bf i} \nu \bar{\sigma}}
d_{{\bf i} \nu \sigma}. \nonumber\\
\label{int}
\end{eqnarray}
It has the intra- and inter-orbital Coulomb interaction
terms as first and second terms, respectively. 
The third and fourth term represents the Hund's coupling and 
the pair hopping. The Hamiltonian possesses only 
two independent interaction parameters due to 
the rotational invariance condition $U = U^{\prime}-2J$. 

Matrix elements of matrices $\hat{\Delta}$ and $\hat{N}$ in Eq. (1) are 
\bea
2\Delta_{\mu\mu} &=& Um_{\mu\mu}+J\sum_{\mu \ne \nu}m_{\nu\nu} \nonumber\\
2\Delta_{\mu\nu} &=& Jm_{\mu\nu}+(U-2J)m_{\nu\mu}
\eea
and 
\bea
2N_{\mu\mu} &=& Un_{\mu\mu}+(2U-5J)\sum_{\mu \ne \nu}n_{\nu\nu} \nonumber\\
2N_{\mu\nu} &=& Jn_{\mu\nu}+(4J-U)n_{\nu\mu},
\eea
where charge densities and magnetizations are given by
\be
n_{\mu\nu} = \sum_{\k \sigma} \langle c^{\dagger}_{\k \mu \sigma}c_{\k \nu \sigma}\rangle,  \,\,\, 
m_{\mu\nu} = \sum_{\k \sigma} \langle c^{\dagger}_{\k \bar{\mu} \sigma}c_{\k \nu \sigma}\rangle .
\ee

\end{document}